\documentstyle[12pt]{article}
\def\hybrid{\topmargin 0pt      \oddsidemargin 0pt
	\headheight 0pt \headsep 0pt
	\textheight 10in         
	\textwidth 6.25in       
	\marginparwidth .875in
	\parskip 5pt plus 1pt   \jot = 1.5ex}

\catcode`\@=11
\def\marginnote#1{}
\newcount\hour
\newcount\minute
\newtoks\amorpm
\hour=\time\divide\hour by60
\minute=\time{\multiply\hour by60 \global\advance\minute by-\hour}
\edef\standardtime{{\ifnum\hour<12 \global\amorpm={am}%
	\else\global\amorpm={pm}\advance\hour by-12 \fi
	\ifnum\hour=0 \hour=12 \fi
	\number\hour:\ifnum\minute<10 0\fi\number\minute\the\amorpm}}
\edef\militarytime{\number\hour:\ifnum\minute<10 0\fi\number\minute}

\def\draftlabel#1{{\@bsphack\if@filesw {\let\thepage\relax
   \xdef\@gtempa{\write\@auxout{\string
      \newlabel{#1}{{\@currentlabel}{\thepage}}}}}\@gtempa
   \if@nobreak \ifvmode\nobreak\fi\fi\fi\@esphack}
	\gdef\@eqnlabel{#1}}
\def\@eqnlabel{}
\def\@vacuum{}
\def\draftmarginnote#1{\marginpar{\raggedright\scriptsize\tt#1}}

\def\draft{\oddsidemargin -.5truein
	\def\@oddfoot{\sl preliminary draft \hfil
	\rm\thepage\hfil\sl\today\quad\militarytime}
	\let\@evenfoot\@oddfoot \overfullrule 3pt
	\let\label=\draftlabel
	\let\marginnote=\draftmarginnote
   \def\@eqnnum{(\theequation)\rlap{\kern\marginparsep\tt\@eqnlabel}%
\global\let\@eqnlabel\@vacuum}  }

\def\numberbysection{\@addtoreset{equation}{section}
	\def\theequation{\thesection.\arabic{equation}}}

\def\underline#1{\relax\ifmmode\@@underline#1\else
	$\@@underline{\hbox{#1}}$\relax\fi}

\def\titlepage{\@restonecolfalse\if@twocolumn\@restonecoltrue\onecolumn
     \else \newpage \fi \thispagestyle{empty}\c@page\z@
	\def\thefootnote{\fnsymbol{footnote}} }

\def\endtitlepage{\if@restonecol\twocolumn \else  \fi
	\def\thefootnote{\arabic{footnote}}
	\setcounter{footnote}{0}}  
\catcode`@=12
\relax

\def\beq{\begin{equation}}
\def\eeq{\end{equation}}
\def\bea{\begin{eqnarray}}
\def\eea{\end{eqnarray}}

\def\nn{\nonumber}

\relax
\hybrid
\begin{document}
\begin{titlepage}
\setcounter{page}{0}
\begin{center}
 \hfill   Landau LITP/3--CP--93 \\
 \hfill   PAR--LPTHE 93--34\\
 \hfill   June 1993\\[.3in]
{\large CRITICAL POINT CORRELATION FUNCTION FOR THE 2D RANDOM
COUPLING ISING MODEL}\\[.2in]

        \large  A.L.Talapov\footnote{E-mail address:
         talapov@itp.sherna.msk.su}\\

         \normalsize {\it Landau Institute for Theoretical
         Physics\\
         GSP-1 117940 Moscow V-334, Russia}\\[.1in]

      	 \large   Vl.S.Dotsenko\footnote{E-mail address:
	 dotsenko@lpthe.jussieu.fr}\\

	 \normalsize {\it LPTHE\/}\footnote{Laboratoire
                              associ\'e No. 280 au CNRS}\\
         {\it Universit\'e Pierre et Marie Curie, PARIS VI\\
	 Tour 16, 1$^{\it er}$ \'etage \\
	 4 place Jussieu\\
	 75252 Paris CEDEX 05, FRANCE\\
     	        and \\
         Landau Institute for Theoretical Physics\\
         Moscow, Russia}

\end{center}

\vskip .2in
\centerline{ ABSTRACT}
\begin{quotation}

High accuracy Monte Carlo simulation results for 1024*1024 Ising system
with ferromagnetic impurity bonds are presented.
Spin-spin correlation function at a critical point
is found to be numerically very close to that of a pure system.
This is not trivial since a critical temperature for the system
with impurities is almost two times lower than pure Ising $T_c$.
Small deviations from the pure behaviour contradict to some of
the competing theories.

\end{quotation}

\end{titlepage}

\newpage

Influence of impurities on a critical behaviour has been a subject of
numerous papers.

For the simplest possible model -- 2D Ising model this problem has been
considered theoretically [1--10],
experimentally \cite{a11,a12} and using computer simulations
[13--18].

Most of the simulations were devoted to thermodynamic properties such as
specific heat, magnetization and magnetic susceptibility. On the other
hand, theories  give direct predictions for spin-spin correlation
function
\beq
\langle S(0)S(r) \rangle
\label{L1}
\eeq
where $r$ is a distance between spins.

We used cluster algorithm special purpose processor (SPP)  \cite{a19,a20}
to get accurate
values of $<S(0)S(r)>$ at a critical point.
The SPP realizes in hardware Wolff cluster algorithm, and therefore does not
suffer of critical slowing down.

The SPP spends $375 ns$ per one cluster spin.
It also has a simple hardware for extremely
fast calculation of spin-spin correlation functions. Time necessary to get
the correlation function for some $r$ is equal to $L^2 * 21 ns$,
where $L$ is a linear lattice size.

We study the following model. Coupling constant $J$ on each bond can take two
positive values: $J_1$ with probability $p$ and $J_0$ with probability $1-p$.
For $p=0.5$ duality relation  \cite{a21} shows that $T_c$ is equal
to that of a pure
model with all horizontal bonds equal to $J_1$ and all vertical bonds equal
to $J_0$. This greatly simplifies simulation data analysis, and so we used
$p=0.5$.

Theoretical models use a small parameter
\beq
g \sim p(J_0 - J_1)^2
\label{L2}
\eeq
This parameter is connected with impurity induced length $l_i$
\beq
\log l_i \sim \frac{1}{g}
\label{L3}
\eeq
To be able to notice deviations from the pure critical
behaviour we should have
\beq
l_i << L
\label{L4}
\eeq
To satisfy this condition we used $L=1024$ and quite different values of $J$:
$J_0 =1$ and $J_1 =0.25$. So in the simulations $g$ is not very small.

Theories deal with continuum limit and infinite lattice size.
Simulations are conducted on a finite lattice with
periodic boundary conditions.
We calculated $<S(0)S(r)>$ for spins, located along one lattice row. In this
case distance $r$ can take only integer values.

Our simulations for the pure case  \cite{a20} showed that
discrete lattice effects
are significant for $r < 8$. Continuum theory can be applied for larger
distances. But the finite size corrections for $r > 8$ are very significant
and should be taken into account explicitly.

Pure Ising correlation function $c_{0}(r)$ for $r/L \rightarrow 0$ has been
calculated in  \cite{a22}
\beq
c_{0}(r)=  \frac{0.70338}{r^{1/4}}
\label{L5}
\eeq
Continuum limit of (1) for the finite lattice with periodic boundary
conditions $c(r,L)$ has been obtained in  \cite{a23}
\beq
c(r,L) \sim \frac{\sum_{\nu =1}^4 \left|\theta _\nu (\frac{r}{2L}) \right|}
{\left|\theta _1 (\frac{r}{L})\right|^{1/4}}
\label{L6}
\eeq
where $\theta$ are Jacobi theta functions.
For our purposes c(r,L) can be written in a simpler form
\beq
c(r,L) \approx A(L) \frac{1+e^{-\pi /4}\left[\sin(\frac{\alpha}{2})+
\cos(\frac{\alpha}{2})\right]
+e^{-9\pi /4}\left[\cos(\frac{3\alpha}{2})-\sin(\frac{3\alpha}{2})\right]}
{\left(\sin(\alpha)-e^{-2\pi}\sin(3\alpha)\right)^{1/4}}
\label{L7}
\eeq
where $\alpha= \pi r/L$.
The coefficient $A(L)$ can be obtained using expression for $c_{0}(r)$.
Formula for $c(r,L)$ is in exellent agreement with simulation
results \cite{a20} for the pure system.

In Fig.1 we show the ratio of computed $<S(0)S(r)>$ to $c(r,L)$ for $L=1024$.

To get mean values of $<S(0)S(r)>$ and standard deviations we used one
thousand samples with different impurities distribution.
For each sample all spins initially were pointing in the
same direction. Two thousand Wolff clusters were flipped to thermalize
spin distribution at critical temperature. Another eight thousand
clusters were flipped to calculate mean values of $<S(0)S(r)>$ for each
sample. For $L=1024$ one cluster flip at $T_c$ requires about $0.1 sec$.

Correlation functions for different $r$ were measured for the same spin
configurations.

Error bars are determined mainly by different behaviour of $<S(0)S(r)>$ for
different impurity distributions
and not by thermal fluctuations for a given sample.

Deviations of $<S(0)S(r)>$ from $c(r,L)$ at $r<8$ are due to the discrete
lattice effects. In fact, at these distances correlation function of the
impure system is extremely close to the pure correlation function. For
$r=1$ difference between them is of the order of $10^{-3}$.
At first sight it seems
to be quite natural, because $r<<l_{i}$ and $<S(0)S(0)>=1$. But the critical
temperatures for pure and impure cases are different, and continuum theory
cannot exclude strong renormalization of $<S(0)S(1)>$.

On the other hand, for $r>8$ pure correlation function practically
coincides with $c(r,L)$. So, Fig.1 shows that impurities decrease
spin-spin correlations at large $r$. Again, this is not trivial because
of the difference of critical temperatures for pure and impure cases.

This result contradicts the renormalization group calculation of DD
\cite{a1} for the spin--spin correlation function, which would instead show
increase if spin correlations, -- see also the discussion in
the first paper of \cite{a13} where the numerical simulation problem has
been defined.

The calculations of DD were based on the interacting fermion model
\cite{a24,a25} representing the 2D Ising model with impurities
\cite{a26}. The replica componets trick has been used, for the fermion
fields renormalization group calculation,
which gave a new critical behavior for the correlation length
and of the specific heat, as compared to the Ising model on a perfect
lattice. The result of \cite{a26} has later been confirmed
by calculations using
other approaches, see [2--7], as well as by the computer simulations [13].
(See however [8,9] which claim a finite specific heat at $T_{c}$).
In these calculations, of the energy operator related quantities,
the replica components number $N$, to be taken eventually equal to $0$,
plays a minor role, that of the diagram counting, both in the original
calculation in [26] using fermions, as well as in other approaches
(bosonization, spin--component replicas)[2--7].

On the other hand, in the calculation of the spin--spin correlation
function in [27] (see also [1]) the fermion replica components
did play a dynamical role, while the other approaches, dealing
more directly with the spin operators, employed replicas for diagram
counting only.

Our present numerical result for the spin--spin function shows that fully
dynamical effects due to replicas of the fermion fields lead
to a wrong result, for this particular problem. The precise reason for this
remains unclear.

The approaches based on bosonization gave the spin--spin function
to be exactly same (in the scaling limit) as that of the pure Ising model.
This is also not supported by our numerical result.

The diviations of $\langle S S \rangle$ in Fig.1 should be compared then
with the standard renormalization group applied directly to the spin
operators. The result of [2,5,6] was that the coefficient of the first order
renormalization group equation vanishes, in the $N \rightarrow 0$ limit
($N$ now is the spin operator replicas). No one had actually calculated
the second order coefficient (let us denote it $C_{2}$), but assuming
it is non--vanishing the result is standard. As the renormalization
group equation one has:
\beq
\frac{d\Gamma (x)}{dx} = - C_{2} g^{2}(x)
\label{L8}
\eeq
Here
\bea
\Gamma (x) = \frac{\langle S(0)S(r)
\rangle}{\langle S(0)S(r) \rangle_{0}}
\label{L9}\\
\langle S(0)S(r) \rangle_{0} \sim c_{0}(r) \sim \frac{1}{r^{1/4}}
\label{L10}\\
x=\log(r)
\label{L11}
\eea
The renormalization group coupling constant $g(x)$ is \cite{a26}:
\beq
g(x)=\frac{g_{0}}{1+\frac{4g_{0}}{\pi}x}
\label{L12}
\eeq
Then from (\ref{L8}) one finds:
\beq
\Gamma (x) = \exp (-\frac{C_{2}g_{0}^{2}x}{1+\frac{4g_{0}}{\pi}x})
\label{L13}
\eeq
and
\beq
\langle S(0)S(r) \rangle \sim \frac{1}{r^{1/4}}
\exp(-\frac{C_{2}g_{0}^{2}x}{1+\frac{4g_{0}}{\pi}x})
\label{L14}
\eeq

So it is not exactly the pure Ising model behavior
as there is a cross--over in the amplitude:
\bea
\Gamma (x) &\approx& 1, \quad \quad \quad \quad \quad \quad  x \ll x_{i}\nn\\
           &\approx& \exp(-\frac{\pi C_{2}}{4}g_{0}), \quad x \gg x_{i}
\label{L15}
\eea
$x_{i} = \pi/ 4g_{0} = \log(l_{i})$ is the impurities driven cross--over
scale.

In the cross--over region $\langle SS \rangle / \langle SS \rangle_{0}
=\Gamma (x) $ is decreasing. This is what we observe in Fig.1.

Suppose that the finite lattice correlation function renormalization is given
by (13) as well. Then we can estimate $C_2$. Value of $g_{0}$, found from the
magnetic sucseptibility, magnetization and specific heat data [13,28], is
somewhere near 0.3. According to Fig.1 $C_2$ turns out to be quite small,
of the order of $10^{-2}$.
This also agrees with the magnetic susceptibility and magnetization
data [28].

More specifically, the estimate by the slope of the plot
in Fig.1 for $r$ relatively small, $8 < r < 32$, where
the finite  size corrections should still be insignificant,
gives $C_{2} \sim 0.03$. Clearly, this is an order of magnitude
estimate. (Here, by finite size corrections,
we mean in fact the small corrections to them due to the extra
factor $\Gamma (x)$, eqs.(13),(14), in the correlation function
of the impure model, as compared to the pure one).

We remark finally that previous measurements of the magnetization in
\cite{a13} has not shown difference with the pure model
because the precision was not sufficient. In fact,
the variation we observe for
$\langle S S \rangle / \langle S S \rangle_{0}$ corresponds,
for small $4g_{0}/\pi x$, to an effective change
in the magnetization index on the value $C_{2}g_{0}^{2}/2$, which is
from $0.125$ to $\sim 0.126$. This is just beyond
the accuracy in \cite{a13}.

\noindent{\large Acknowledgements}

We acknowledge useful discussions with V.B.Andreichenko,
Vik.S.Dotsenko, \\
V.L.Pokrovsky, W.Selke, and L.N.Shchur.

\newpage



\newpage

{\bf Figure Captions}

\vspace {1.5truecm}

Fig.1. Ratio of $<S(0)S(r)$ to $c(r,L)$. Solid line connects data points
for the system with impurities at $(1/T_c)=.8070519$.
Dashed line shows data for the pure Ising
model \cite{a23} at $(1/T_c)=.4406868$.

\end{document}